\documentclass[final,3p,times,onecolumn,authoryear]{elsarticle}

\usepackage{amssymb}
\usepackage{hyperref}

\usepackage{lineno}
\usepackage{color}
\usepackage{ulem}

\def\tb{\textcolor{black}}

\journal{Chemical Engineering Science}

\begin{document}

\begin{frontmatter}




\title{The role of viscosity contrast on plume structure in laboratory modeling of mantle convection}



\author[jncasr]{Vivek N. Prakash\fnref{stan}}
\fntext[stan]{\\ \textsuperscript{1}Present address: Department of Bioengineering, Stanford University, California, USA}

\author[jncasr]{K. R. Sreenivas\corref{cor1}}
 \cortext[cor1]{Corresponding author, Email address: krs@jncasr.ac.in}
\address[jncasr]{Engineering Mechanics Unit, Jawaharlal Nehru Centre for Advanced Scientific Research, Jakkur P.O., Bangalore - 560064, India.}

\author[iisc]{Jaywant H. Arakeri}
\address[iisc]{Department of Mechanical Engineering, Indian Institute of Science, Bangalore - 560012, India.}

\begin{abstract}
We have conducted laboratory experiments to model important aspects of plumes in mantle convection. We focus on the role of the viscosity ratio $U$ (between the ambient fluid and the plume fluid) in determining the plume structure and dynamics. 
We build on previous studies to highlight the role of viscosity contrast in determining the morphology of mantle plumes and provide detailed visualisations and quantitative information on the convection phenomenon. In our experiments, we are able to capture geophysical convection regimes relevant to mantle convection both for hot spots  (when $U > 1$) and plate-subduction (when $U < 1$) regimes. The planar laser induced fluorescence (PLIF)  technique is used for flow visualization and characterizing the plume structures. The convection is driven by compositional buoyancy generated by the perfusion of lighter fluid across a permeable mesh and the viscosity ratio $U$  is systematically varied over a range from 1/300 to 2500. The planform, near the bottom boundary for $U=1$, exhibits a well-known dendritic line plume structure. As the value of $U$ is increased, a progressive morphological transition is observed from the dendritic-plume structure to discrete spherical plumes, accompanied with thickening of the plumes and an increase in the plume spacing. In the vertical section, mushroom-shaped plume heads at $U=1$ change into intermittent spherical-blob shaped plumes at high $U$, resembling mantle plume hot spots in mantle convection. In contrast, for low values of $U (~1/300)$, the regime corresponds to subduction of plates in the mantle. In this regime, we observe for the first time that plumes arise from a thick boundary with cellular structure and develop into sheet-plumes. We use experimental data to quantify these morphological changes and mixing dynamics of the plumes at different regimes of $U$. We also compare our observations on plume spacing with various models reported in the literature by varying the viscosity ratio and the buoyancy flux. 

\end{abstract}


\end{frontmatter}



\section{\label{sec:level1}Introduction\protect\\ }

The study of convection is important, with various parametric regimes of convection being relevant in different fields, for example, high Rayleigh number convection is important both in natural processes (e.g. atmospheric and mantle convection) and in engineering applications (e.g. \tb{chemical engineering industry} and metallurgy) and has been studied extensively~(\cite{Ahlers2009}). Mantle convection occurs at moderately high Rayleigh and Prandtl numbers ($Ra\approx 10^6$ to $10^8$ and $Pr\approx10^{24}$) in a configuration similar to Rayleigh-Benard Convection. Mantle convection in the earth is an important process by which heat is transported from the core to the surface and is responsible for volcanism, plate tectonics and orography (see reviews:~\cite{Humphreys2011,Ribebook2007,Jellinek2004}). Morgan~(\cite{Morgan1971}) put forward the hypothesis that `mantle plumes' are responsible for the origin of `hotspots' on the earth. Mantle plumes detach from the thermal boundary layer (being lighter and less viscous due to higher temperature) at the core-mantle boundary and rise in a more viscous ambient mantle. These mantle plumes are difficult to observe and available information about them is based on indirect geological measurements~(\cite{Zhao2001}), and analog laboratory experiments like ours and numerical simulations~(\cite{Kellogg1997,Vankeken1997}).
The mantle primarily consists of solid silicate rock which can be regarded to behave as a fluid at geological timescales with a high viscosity $\approx10^{18} m^2s^{-1}$. Unlike Rayleigh-Benard convection, in mantle convection there are large variations in viscosity, pressure and composition. Capturing all these parameters in a single experiment is a challenge. The viscosity of the mantle is dependent on the composition, pressure and temperature~(\cite{Schubertbook}). A temperature increase of $100\,^{\circ}\mathrm{C} - 300\,^{\circ}\mathrm{C}$ can reduce the viscosity of the mantle by a factor of $100$ to $1000$ (see~\cite{Daviesbook}). Similarly, subduction - a geological process in which one edge of crustal plate is forced below its neighboring crustal plate, and its initiation and sustainability is also a topic of recent studies~(\cite{Ueda2008,Sizova2010,RegenauerLieb2000,Solomatov2004,Schubertbook}). Subduction induced by cooler, heavier oceanic crust that plunges into the mantle~(\cite{Ueda2008,Sizova2010}), corresponds to the situation where a more viscous plume is moving into low viscosity mantle driven by a density difference. Thus, the prime factors that set apart mantle convection are the extremely high value of Prandtl number and high viscosity contrast between plumes and the ambient fluid. These two factors play a crucial role in determining the plume longevity, mixing, rise velocities and hence, the heat transfer within the mantle~(\cite{Olson1985,Vankeken1997,Lenardic2009}). 

In this work, we investigate the role of viscosity ratio $U$ (viscosity of ambient fluid / viscosity of plume fluid) on the structure and dynamics of plumes. In our experiments, we have used compositional buoyancy to drive the convective flow. In this study, we are concerned with the effect of the viscosity ratio, $U$, on the (a) spacing and morphology of plumes, (b) structure and dynamics of the rising plumes and (c) longevity and mixing of plume with the ambient fluid. 


In previous convection experiments driven by thermal buoyancy for example~(\cite{Manga1999,Lithgow2001}), the buoyancy flux and fluid viscosity were coupled as the fluids had a temperature-dependent viscosity. The use of compositional buoyancy to drive the convection provides us the opportunity to study independently the effect of viscosity contrast on the plume structure decoupled from other parameters. Compositional buoyancy has been used to study plumes rising from point sources at different viscosity ratios~(\cite{Whitehead1975,Olson1985}). These studies have led to the standard accepted model of a mantle plume: a large, bulbous head  trailed by a narrow conduit or tail connecting it with its source. In the present study, plumes arise as boundary layer instabilities from the bottom surface, driven by compositional buoyancy. The fluid dynamical regime corresponds to something in between the flow associated with a Rayleigh-Taylor instability and Rayleigh-Benard convection. Previous experiments on convection across a mesh, have reported different regimes depending on the magnitude of the concentration differences across the mesh, a diffusive regime~(\cite{Baburaj2005}) similar to Rayleigh-Benard convection, and an advection regime~(\cite{Baburaj2008}) with the existence of a  through flow across the mesh. In our experiments, we externally impose a through flow across the mesh, and the advection velocities are $\sim0.083 cm s^{-1}$ (at least 10 times greater than the previous work by~\cite{Baburaj2008}). The experiments of ~\cite{Jellinek1999}) used a similar setup but their primary motivation was to study mixing in different viscosity ratio regimes ($U \approx 1/850$ to $20100$). 

Our focus is to study the effect of $U$ on plume structure and plume dynamics by flow visualization experiments. We report results on the planform plume structures and quantify the changes in plume morphology, plume dynamics and the plume mixing effectiveness over a wide range of viscosity ratios ($U \approx 1/300$ to $2500$) using image processing. 
Here, since we have used concentration differences to provide compositional buoyancy, the Schmidt number (Sc) is a proxy to the Prandtl number. In our experiments, we are able to simultaneously achieve high Rayleigh numbers $\approx10^{11}$ and high Schmidt numbers $\approx 10^6$. 

\tb{In addition to mantle convection, this work is also of interest to chemical engineers because the viscosity contrast as a parameter is relevant in the chemical process industry, e.g. in blending of additives into polymer melts. The new mixing effectiveness measure we propose here provides a useful and stringent tool to quantify mixing in a variety of industrial contexts, e.g. in batch versus continuous mixing in various chemical processes. }

In section~\ref{sec:exp}, we present details of the experimental setup and the methodology, followed by the results in section~\ref{sec:res} with flow visualization pictures of plume structures showing the dependence of the convection pattern and dynamics on the viscosity ratio $U$, $Ra$ and $Sc$ numbers with a constant buoyancy flux. We also report preliminary results on the effect of varying the buoyancy flux while $U$ is held constant. In the concluding section~\ref{sec:con}, we summarize the results from the present study.

\section{Experimental Set up and Methodology}\label{sec:exp}

\begin{figure*}
\centering
\includegraphics[width=1\textwidth]{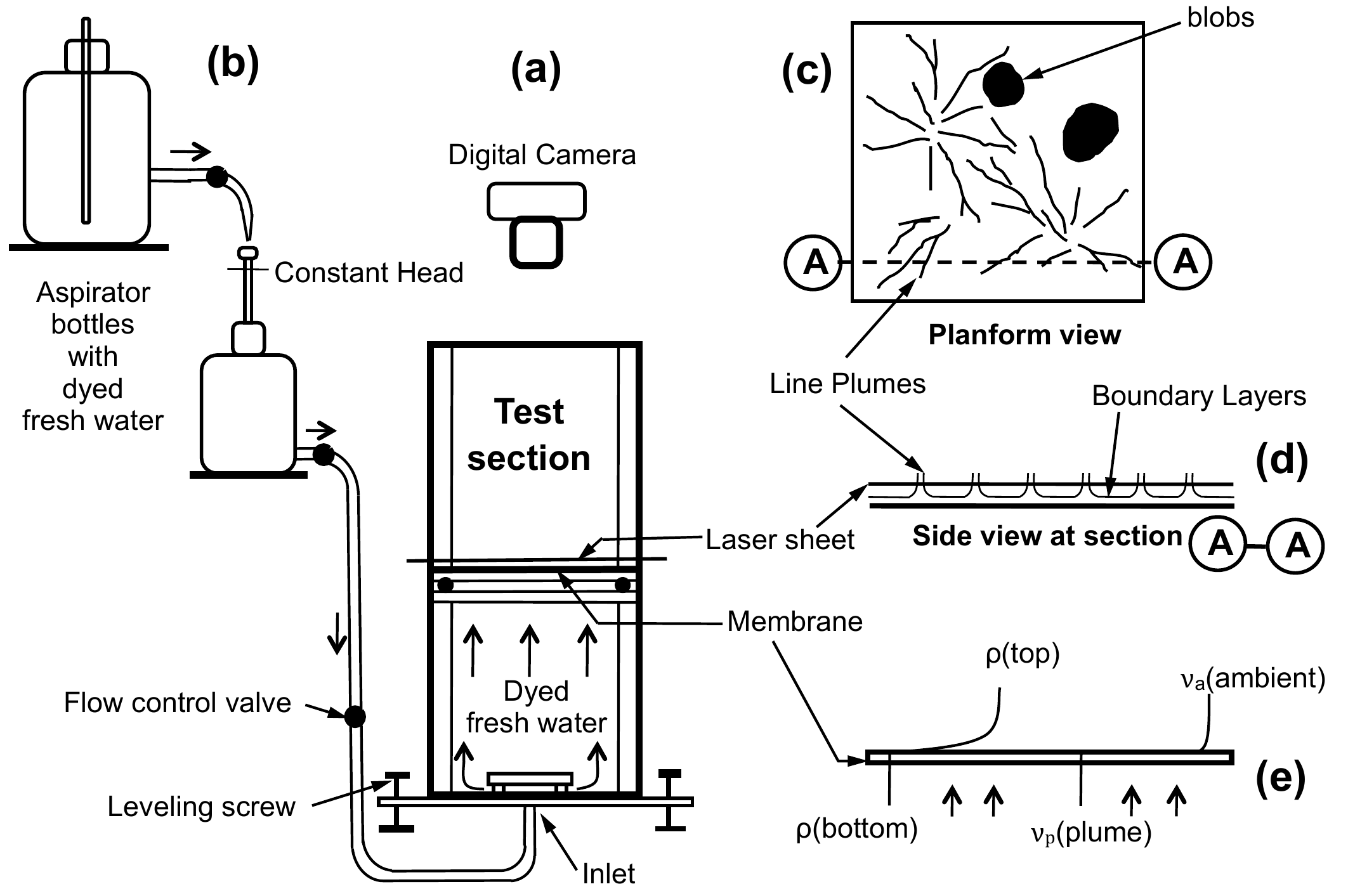}
\caption{Schematic of the experimental setup, (a) main experimental setup with test-section and digital camera, (b) constant flow-rate arrangement, (c) typical planform view of the plume structure just above the mesh, obtained by passing a horizontal laser-sheet above the mesh and dyed fluid from the lower chamber, (d) arrangement in the vertical section (side-view) and (e) density and viscosity profiles across the mesh.} 
\label{fig:exptsetup}
\end{figure*}

A schematic of the overall experimental setup and visualization process is shown in Figure~\ref{fig:exptsetup}. The experiments have been conducted in a square cross-section tank that is divided into two chambers by a permeable mesh. The convection is driven by a concentration difference across the mesh with heavier sugar solution in the upper chamber and lighter fresh water in the bottom chamber. The test section is the upper chamber of the tank which has a $15.5 cm \times 15.5 cm$ cross-section and extends to a height of $26 cm$. A constant-head flow delivery arrangement~(\cite{Arakeri2000}) is used to perfuse lighter water into the setup from the bottom chamber across the mesh, at a constant flow rate. This experimental design enables the study of plumes rising from the boundary layer on the mesh, over a range of viscosity ratios.

The primary source of data in our experiments is from still-images and videos acquired from flow visualization using the Planar-Laser-Induced-Fluorescence (PLIF) technique. The beam from a 50 mW, 532 nm diode laser is converted into a planar light sheet using a plano-concave lens. A horizontal planar laser sheet is used to visualize the planform structures and a vertical planar light sheet is used to record vertical sections of plumes. A small amount ($\approx 0.5$ppm) of fluorescent dye (Rhodamine-6G) is dissolved in the bottom chamber fluid (plume fluid), which acts as a passive tracer and helps to visualize the buoyant plume structures. Still Images were captured at intervals of 5 seconds using a Digital SLR camera (Canon EOS 350D) and videos were recorded using either a Nikon D90 Digital SLR camera (video mode) or a Handycam (Sony DCR-DVD201E). 

Two different meshes have been used in our experiments in different conditions: a nylon mesh (pore size $29\mu$$m$, open area factor - 0.2) and a steel mesh (pore size $427\mu$$m$, open area factor - 0.45). The nylon mesh was used in experiments where $U = 1$ and $U\gg1$. \tb{However, it is difficult to force viscous fluids through this nylon mesh due to its small pore size, and also since nylon is a soft material, it would bulge and eventually tear. In order to overcome this limitation, we used a steel mesh in our experiments when $U\ll1$, as it has a larger pore size and is rigid. The effect of the mesh pore size in our experiments is expected to be negligible since it has been chosen to be sufficiently small ($< 0.3 \%$) compared to the dimensions of the test section.} 

Different values of $U$ have been achieved in the present experiments by enhancing the viscosity of either the higher density upper chamber fluid (for cases $U > 1$) or the bottom chamber fluid (for cases $U < 1$) using Sodium-CarboxyMethylCellulose (ÔAqualonÕ 7H4F grade CMC, Hercules Inc). Addition of small amounts (max $1\%$ by weight) of CMC enables the variation of viscosity by three orders of magnitude, while causing negligible effects on other fluid properties such as density~(\cite{Tait1989,Davaille1999}). The CMC solutions are known to be Newtonian at the shear rates encountered ($< 1 s^{-1}$) in our experiments. We have measured the absolute viscosity of CMC solutions (in $mPas$) versus the concentration of CMC in solution ($\%$ weight) using a rheometer (Rheolyst series AR1000, TA Instruments)  at shear rates $\approx1 s^{-1}$ (for more details see~\cite{VivekMS}).

The main parameters governing the flow are as follows: compositional Rayleigh number ($Ra_c$), flux Rayleigh number is $Ra_f$, Schmidt number ($Sc$) is the proxy for Prandtl number, $U$ is the viscosity ratio, $B$ is  the buoyancy flux at the mesh boundary, $Re_a$ is the Reynolds number based on Deardorff velocity scale, and $Re_{mes,a}$ is the Reynolds number based on the measured velocity scales:
 \begin{equation}
Ra_c= \frac {g \Delta\rho H^3} {\rho \nu_{a} \alpha_{m}}
\end{equation}
\begin{equation}
Ra_f= \frac {g \Delta\rho v_m H^4} {\rho \nu_{a} \alpha_{m}^2} = Ra_c \Big( \frac {v_m H} {\alpha_{m}}\Big)
\end{equation}
\begin{equation}
Sc= \frac {\nu_{a}} {\alpha_{m}}
\end{equation}
\begin{equation}
U= \frac {\nu_{a}} {\nu_{p}} 
\end{equation}
\begin{equation}
B= g \frac {\Delta\rho} {\rho} v_m + g \frac {\Delta\rho} {\rho} \frac {\alpha_{m}} {\delta} \simeq g \frac {\Delta\rho} {\rho} v_m
\label{eq:fluxeqn1}
\end{equation}
\begin{equation}
 Re_a = \frac {wH} {\nu_{a}}
 \end{equation}
 \label{eq:Re}
 \begin{equation}
 Re_{mes,a} = \frac {v_{mes} H} {\nu_{a}}
 \end{equation}
 \label{eq:Remes}
where, $g$ is the acceleration due to gravity, $H$ is the height of the fluid layer in the upper chamber, $\Delta\rho$ is the density difference between the ambient fluid and the plume fluid, $\rho$ is the density of the plume fluid, $\alpha_{m} (= 5\times10^{-10} m^2s^{-1})$ is the mass diffusivity of sugar in water, $\nu_{a}$ is the kinematic viscosity of the ambient (upper tank) fluid, $\nu_{p}$ is the kinematic viscosity of the plume (lower tank) fluid, $v_m$ is the through flow velocity of the injected plume fluid across the mesh, and $\delta$ is the thickness of the concentration boundary layer above the mesh, $w$ is the Deardorff velocity scale~(\cite{Deardorff1970}): $w = (BH)^{1/3} = (g\frac {\Delta\rho} {\rho} v_{m}H)^{1/3}$. As we shall see below, $w$ is valid only for $U = 1$. $v_{mes}$ is the experimentally measured vertical plume velocity (see section $3.2$). Note that the Rayleigh numbers ($Ra_c$ and $Ra_f$) and the Reynolds numbers ($Re_a$ and $Re_{mes,a}$) are based on the kinematic viscosity of the ambient fluid. In the present experiments, the ratio of dynamic viscosities is nearly same as the ratio of the kinematic viscosities, as the density difference is very small. \tb{Among the governing parameters defined above (Eqs.~[1-7]), $Ra_c$, $Sc$ and $U$ are independent parameters.}

In the present study, $U = 1$ (no viscosity contrast) serves as the base case, because this regime is similar to the well-studied system of turbulent natural convection. From the convective stability criterion in Rayleigh-Benard convection, we estimate $\delta$ as, $\delta \sim 10(\nu D / g \frac {\Delta\rho} {\rho})^{1/3}$, which is $\sim1mm$ in our experiments. The buoyancy flux $B$  defined above (in eq~\ref{eq:fluxeqn1}), is the sum of the advective flux and the diffusive flux. In our experiments, an estimate of the relative importance of the two terms reveals that the diffusion term is at the most (1/500) of the advective term (the ratio varies from 1/500 to 1/5000 in our experiments). Hence, in this work, the buoyancy flux is dominated by the advection term $B= g \frac {\Delta\rho} {\rho} v_m$, and the diffusion flux term is negligible. The diffusion term is significant when there is no externally imposed through-flow across the mesh (see for e.g.~\cite{Baburaj2005}). Here, the Peclet number, a ratio of the advection to diffusion, $Pe = (v_m \delta / \alpha_{m}) \sim 500$ for the $U=1$ case, indicating that our experiments are clearly dominated by advection. Although the convection dynamics near the mesh are dominated by advection, we assume that the dynamics in the bulk would be similar to high $Ra$ number turbulent natural convection over horizontal surfaces.

\begin{table}
\caption{A summary of parameters covered in the different sets of  experiments. $U$ is the viscosity ratio, CMC $\% wt$ is the amount of Sodium-CarboxyMethylCellulose added by weight \% in solution, $\Delta\rho$ is the density difference, $Ra_c$ is the compositional Rayleigh number, $Ra_f$ is the flux Rayleigh number, $Sc$ is the Schmidt number, $Re_a$ is the Reynolds number based on Deardorff velocity scale, $Re_{mes,a}$ is the Reynolds number based on the measured velocity scales and have been provided only for cases where velocities were measured, and $v_m$ is the through flow velocity of the injected plume fluid across the mesh.}
\footnotesize{
\begin{center}
 \begin{tabular}{c|c|c|c|c|c|c|c|c|c|c}
 \hline 
       \bf{Expt.}&\bf{Viscosity}&\bf{CMC}&\bf{CMC}&$\Delta\rho$&\bf{$Ra_c$}&\bf{$Ra_f$}&\bf{$Sc$}&\bf{$Re_a$}&\bf{$Re_{mes,a}$}&\bf{$v_m$}\\ 
       \bf{\#}&\bf{Ratio, U}&$\% wt$&addition&$kg m^{-3}$&$\times10^8$&$\times10^{13}$ &$\times10^3$& & &$(cm  s^{-1})$ \\
\hline 
             1&1&0&--&2.4&2934&4522&2&2230&1385&$0.083$\\
             2&3&0.025&tank fluid&2.4&978&1507&6&744&--&$0.083$\\
             3&5&0.05&tank fluid&2.4&586&904&10&446&--&$0.083$\\
             4&7&0.1&tank fluid&2.4&419&646&14&319&--&$0.083$\\
             5&15&0.15&tank fluid&2.4&195&301&30&149&--&$0.083$\\
             6&25&0.2&tank fluid&2.4&117&181&50&89&38&$0.083$\\
             7&65&0.3&tank fluid&2.4&45&70&131&34&--&$0.083$\\
             8&130&0.4&tank fluid&2.4&22&35&262&17&--&$0.083$\\
             9&300&0.5&tank fluid&2.4&9.78&15&604&7.4&1.6&$0.083$\\
             10&470&0.6&tank fluid&2.4&6.24&9.6&946&4.7&--&$0.083$\\
             11&930&0.75&tank fluid&2.4&3.15&4.8&1873&2.4&--&$0.083$\\
             12&2500&1&tank fluid&2.4&1.17&1.8&5035&0.9&0.1&$0.083$\\
             \hline
             13&1/300&0.5&input fluid&2.4&2934&981&2&1340&338&$0.018$\\
             14&1/130&0.4&input fluid&2.4&2934&1580&2&1572&354&$0.029$\\
             15&1/65&0.3&input fluid&2.4&2934&1852&2&1657&662&$0.034$\\
             16&1/25&0.2&input fluid&2.4&2934&4522&2&2230&--&$0.083$\\
             \hline
             17&1&0&--&2.4&2934&1090&2&1389&--&$0.020$\\
             18&1&0&--&2.4&2934&2234&2&1764&--&$0.041$\\
             19&1&0&--&2.4&2934&3378&2&2024&--&$0.062$\\
             20&1&0&--&2.4&2934&4522&2&2230&--&$0.083$\\
             \hline
             21&1/300&0.5&input fluid&4.8&5868&1960&2&1690&--&$0.018$\\
    \end{tabular}
    \end{center} 
    \label{tab:expparam}
    }
\end{table}

In this work, we focus on studying the plume structure by changing $U$ systematically at a constant $B$. Table~\ref{tab:expparam} summarizes the parameter ranges covered in the different sets of experiments that have been conducted in the present study. The parameter ranges we have covered are: $U \approx$ $1/300$ - $2500$, $Ra_{c} \approx 10^8 -10^{11}$ and $Sc \approx 10^3 - 10^6$. In experiment sets 1 to 12, the initial density difference was $\Delta\rho$ = 2.4 kg $m^{-3}$ and the volume flow rate of the injected plume fluid is held constant throughout the duration of the experiment at 4 mL $s^{-1}$. This volume flow rate corresponds to a \tb{through flow velocity $v_{m} = 0.083$ cm $s^{-1}$ just above the mesh,} which is small ($\approx1/10$) compared to the typical plume rise velocities. In experiments where $U < 1$ (set 13, 14, 15) there were limitations in maintaining a constant through flow velocity across the mesh, as viscous fluid had to be forced through the mesh. Hence, the through flow velocities quoted are averaged over the duration of the experiment. However, in experiment sets 13--16, the initial density difference was still maintained constant at $\Delta\rho$ = 2.4 kg $m^{-3}$.

The buoyancy flux $B$ is constant in experiments 1--12 and 16 (in sets 13, 14 and 15, B is not constant due to limitations in maintaining a constant flow through the mesh). In experiment sets 17--20 $\&$ set 21 (see Table~\ref{tab:expparam}), we present some preliminary results with varying $B$ at fixed $U$. In sets 17--20, we vary the through flow velocities of the injected plume fluid ($v_m$) (by changing the volume flow rates) and in set 21, we double the initial density difference to $\Delta\rho$ = 4.8 kg $m^{-3}$. The flow across the mesh is always upwards; i.e. from the bottom chamber into the top chamber. A typical experiment would last for about $20$ minutes. 


\tb{In our experiments, when $U = 1$, we see that the plume structure planforms are almost the same (in a statistical sense) at different times; we consider this situation to be a `quasi-steady state'. When $U>1$ and $U < 1$, there are initial transients which take time to settle down (a few minutes) into the `quasi-steady state' - where the planforms look the same over time in a statistical sense (see supplementary material for more details). In all our analysis, we have considered data during the quasi-steady phase.  }


\section{\label{sec:res}Results and Analysis}

\subsection{Near-base planform plume structure} \label{sec:planform}

\begin{figure}
\centering
\includegraphics[width=1.05\textwidth]{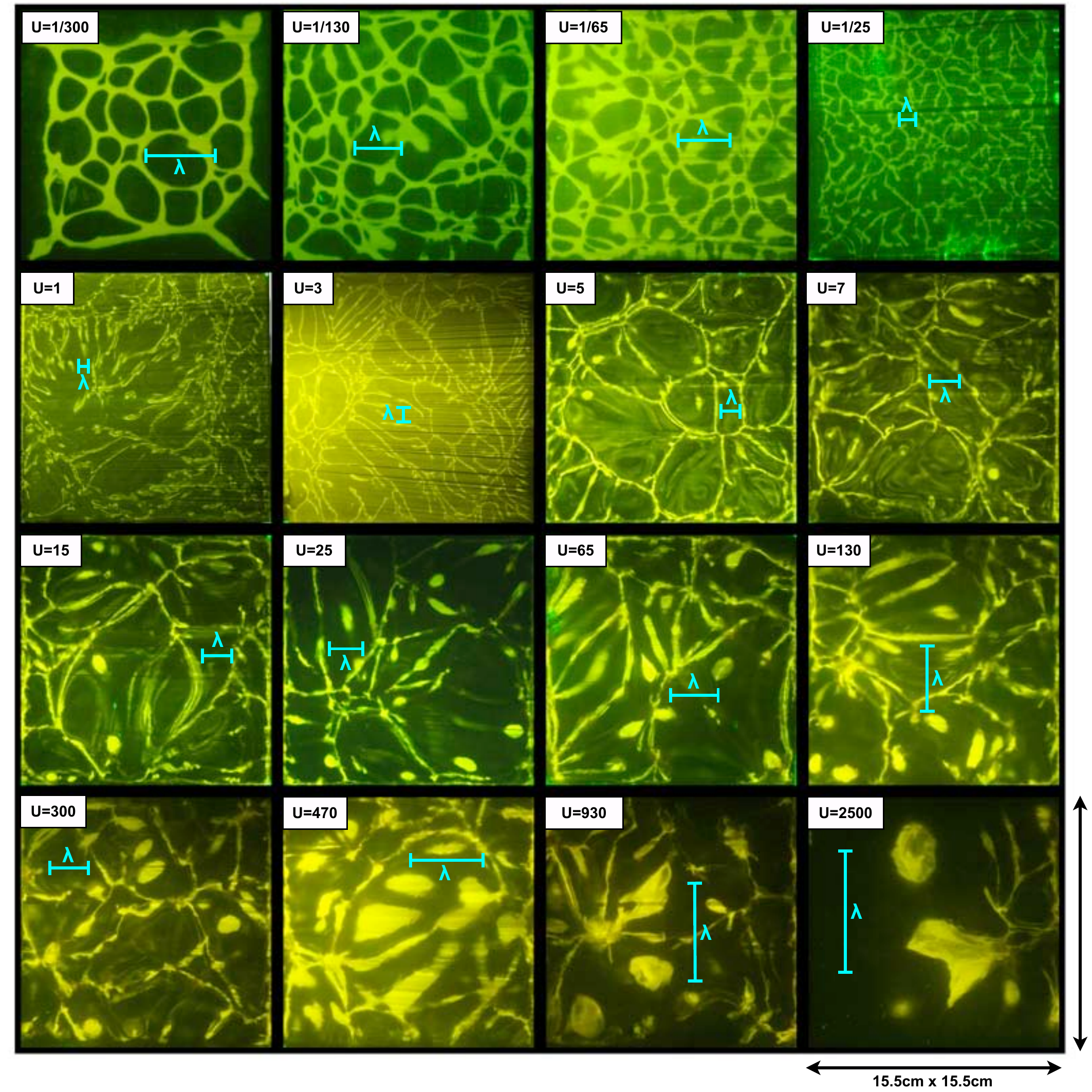}
\caption{ {\bf{Montage of near-wall plan view images of plume structures.}} 
The raw images from the experiment are shown. The regions in yellow/green colour are plumes and the dark background is the ambient fluid. The viscosity ratio $U$ progressively increases from top to bottom $(U = 1/300, 1/130, 1/65, 1/25, 1, 3, 5, 7, 15, 25, 65, 130, 300, 470, 930, 2500)$ as shown. The characteristic plume spacing $\lambda$ (from the analysis) is indicated in each case in cyan color. }
\label{fig:planformmontage}
\end{figure}

In Figure~\ref{fig:planformmontage}, a montage of the raw images ($2052\times2052$ pixels, covering test section of an area $15.5 cm\times15.5 cm$ with a resolution of $0.075$ mm per pixel) is shown. The images in Figure~\ref{fig:planformmontage}  show the characteristic planforms in the range of $U$ covered in the present study; the corresponding $U$ value is indicated in each image. In these experiments, only the viscosity ratio $U$ is varied and other parameters are held constant (flow rate = 4 mL $s^{-1}$, density difference $\Delta\rho$ = 2.4 kg $m^{-3}$). In general, the gravitational instability of the lighter fluid gives rise to plumes that arise from the boundary layer on the mesh. The planforms are the horizontal cross-sections of the rising plumes, as seen from the top, and provide information on the dynamics of plume formation and their eventual detachment from the boundary layer. Also, since the plumes form in the the near-wall region, the spacing and thickness can be studied in this region as a function of $U$. The images have been selected at a quasi-steady state in the experiment. A repetition of some experiments showed that the planforms have the same characteristics in an average sense, so the selected images are representative of the corresponding case of $U$.  
In experiments where $U > 1$ (lower plume-fluid viscosity), the planforms are just above the mesh ($\approx 1mm$). However, in experiments where $U < 1$ (higher plume-fluid viscosity), we observed a thicker boundary layer ($\approx 2 - 4 mm$ thickness) above the mesh with the formation of cellular patterns just above the thick boundary layer. Here, we select images of the planform structures just above the boundary layer. 

We first provide a qualitative description of changes in planform structures with varying $U$. Here, the $U = 1$ (no viscosity contrast) experiment serves as the base case (see Figure~\ref{fig:planformmontage}, $U=1$). The planform mainly consists of line plumes with a dendritic structure and is a familiar structure observed in previous experimental studies~(\cite{Husar1968,Theerthan1998,Baburaj2005}) for high $Ra$ turbulent natural convection. As $U$ increases (increasing viscosity of the ambient fluid), the spacing between the line plumes as well as their thickness increase (Figure~\ref{fig:planformmontage}, Images with $U = 3 ,5$ $\&$ $7$). Further increasing $U$ beyond 15 (Figure~\ref{fig:planformmontage}), we observe a transition in the planform structure from line plumes to discrete blobs, which start to dominate the morphology. Whereas line plumes become thicker and gradually lose their prominence by $U = 300$ (Figure~\ref{fig:planformmontage}), and blobs become the characteristic structures. As $U$ is increased further, size of the blobs increase. Finally, at $U = 2500$ (Figure~\ref{fig:planformmontage}), only 2-3 isolated large blobs are seen. The slenderness ratio $(w/l)$ of the plumes in the planform tends to unity as $U$ is increased from $1$ to $2500$. Here, we define $w$ as the shorter dimension - `width' or `thickness' of the plumes, and $l$ as the longer dimension - lengths of the plume cross-sections in the planform. 

In the other end of the regime, where $U < 1$ (bottom chamber fluid being more viscous), a cellular structure is observed.  When $U = 1/300$ (Figure~\ref{fig:planformmontage}), the plumes develop over a thick boundary and are connected to form approximately elliptical cells. The cell size decreases as the value of $U$ is increased towards 1 (refer to Figure~\ref{fig:planformmontage} till $U = 1/65$). When $U$ is 1/25, the cells are no longer able to form the closed cellular structure. Still the plumes for $U=1/25$ are thicker and have a more uniform size than that for $U=1$, and are closely packed with smaller spacing. In summary, when $U\gg1$, corresponding to hot mantle plumes, we observe discrete blobs with larger spacing. When $U\ll1$, corresponding to the subduction scenario, cellular structures form over a thick boundary layer. Line plumes with a dendritic structure are observed when U is equal to 1. Some of these experimental results have also been captured in numerical simulations (\cite{Tomar2015}). 

\tb{The effect of walls on the plume spacing is an interesting question, and we will briefly discuss our observations here. When $U = 1$, we have observed large scale flows, similar to the large scale convection (LSC) rolls found in classical Rayleigh-Benard Convection. Here, it is well known that walls do not affect the small scale plume structure.
However, in our experiments when $U > 1$ or $U < 1$, the number of plumes produced are less, and we see less large scale flows. When $U > 1$, the plumes frequently arise from a given location on the mesh once they start forming, and these locations (spacings) and their formation can probably be attributed to wall effects. In the extreme case, when $U = 2500$, there are very few plumes produced ($\sim$ 2-3) and the plume spacing may be affected due to wall effects. When $U << 1$, the planform plumes show cellular structures with a well-defined boundary (see planform corresponding to U = 1/300 in Figure~\ref{fig:planformmontage}), we believe these are also due to wall effects. }

\begin{figure}
\centering
\includegraphics[width=0.98\textwidth]{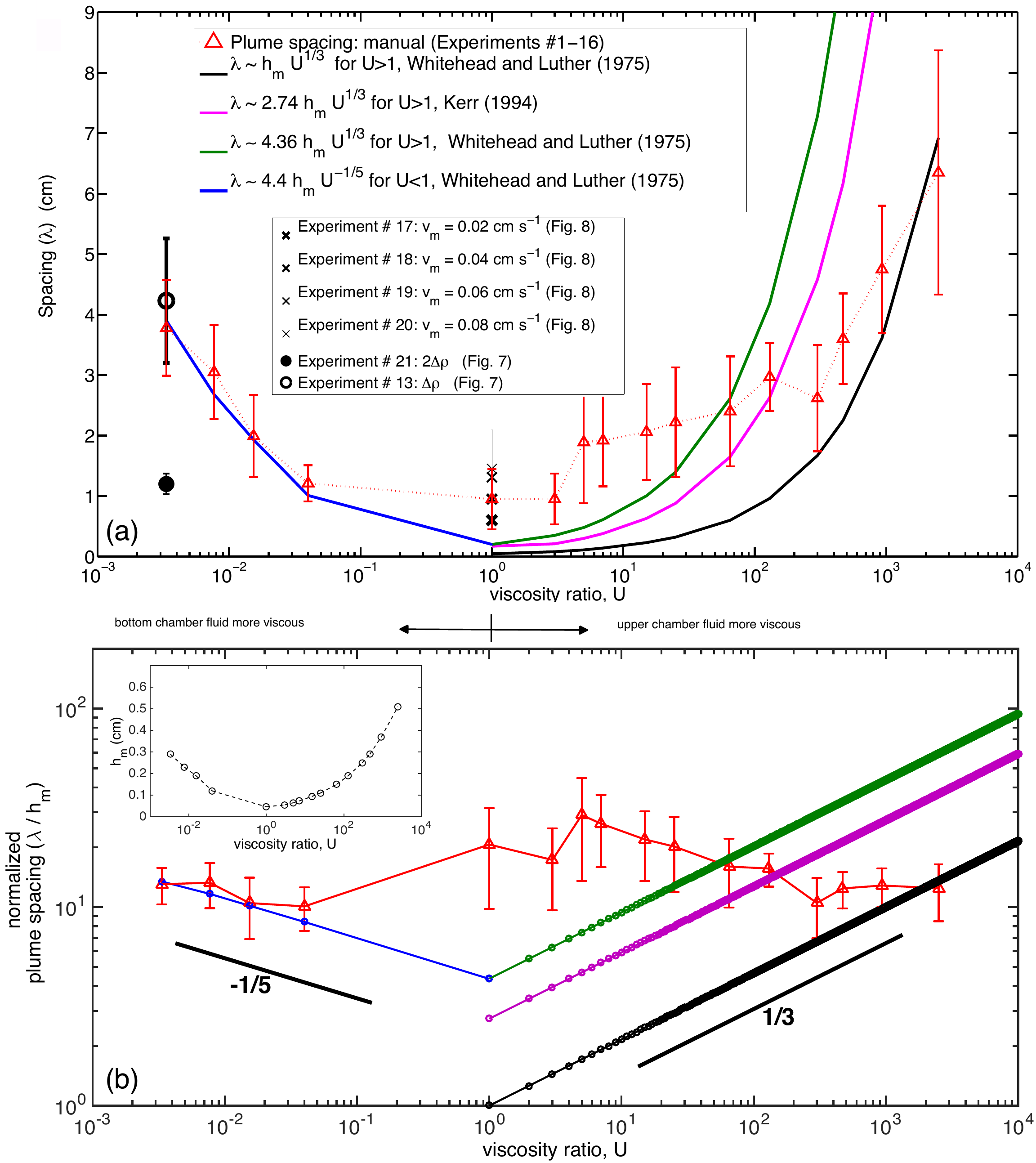}
\caption{The variation of plume spacing $\lambda$ with viscosity ratio, $U$. (a) Here, we compare our plume spacing results (in $cm$) from the manual inspection method with predictions of plume spacing $\lambda$ from previous theories~(\cite{Kerr1994,Whitehead1975}) (shown by lines of different colours). Here, we also plot the spacing values from experiment sets ($17-20, 13, 21$) where the buoyancy flux was varied (represented by black color data points). (b) The normalized plume spacing, i.e. $\lambda / h_{m}$ versus $U$. Inset: variation of $h_m$ with $U$. The legend for (b) is the same as in (a). } 
\label{fig:spacingthickness}
\end{figure}

We have seen that the planform plume structures at different $U$ have random orientations and locations, along with differences in morphology. Quantifying the plume spacing (hereafter referred to as $\lambda$) from these images (Figure~\ref{fig:planformmontage}) is hence a non-trivial task. We have adopted an autocorrelation-based image analysis procedure to quantify the plume spacing  and plume thickness (more details on the procedure can be found in~\cite{VivekMS} and supplementary material). 

Here, we have chosen to show plume spacing results from a manual inspection method. We have determined the plume spacing by careful manual mouse-clicks on the plumes in the different binary images in Matlab, and then take the value averaged over a number samples to obtain $\lambda$.  These results are shown in Figure~\ref{fig:spacingthickness}(a). We have observed that both the manual inspection and autocorrelation methods capture the same trend (see supplementary material): the plume spacing is a minimum at $U = 1$, and increases as $U$ moves away from one (Figure~\ref{fig:spacingthickness}(a)). 

In the literature on high $Ra$ convection, theoretical predictions of plume spacing as a function of $Ra$ have been made. For the general case of convection over heated horizontal surfaces, the plume spacing, $\lambda$, based on stability arguments for $Pr \sim 1$, is predicted to be~(\cite{Theerthan1998}):
\begin{equation}
Ra_{\lambda}^{1/3} = \lambda /Z_{w} = 53
\label{eq:Theerthan}
\end{equation}
where $Z_{w}$ is a near-wall length scale for turbulent convection. When the convection is driven by density differences across a mesh, in the diffusion regime, similar to Rayleigh-Benard convection, the plume spacing for $Pr > 1$ (Schmidt number being proxy to Pr) has been shown to be~(\cite{Baburaj2005}): 
\begin{equation}
Ra_{\lambda}^{1/3} = \lambda /Z_{w} = 92
\label{eq:Baburaj}
\end{equation}
It must be noted that these theoretical expressions for turbulent convection are valid for the case of $U=1$, and when the flux across the mesh is purely by diffusion. In our experiments, for $U=1$, the estimated plume spacing would be $0.15 cm$ (according to Eq~\ref{eq:Theerthan} from ~(\cite{Theerthan1998}) and $0.21 cm$ (according to Eq~\ref{eq:Baburaj} from~\cite{Baburaj2005}), compared to the value of $0.95 cm$ from our measurement. The disagreement is to be expected due to the externally imposed through flow across the mesh in our experiments. This result will be discussed further in section~\ref{sec:varyingflux}, where we study the effect of the through flow velocity on the plume spacing.

In Figure~\ref{fig:spacingthickness}, we compare our plume spacing results at $U>1$, to a previous study of melting driven by compositional convection, by~\cite{Kerr1994}. Using a scaling analysis, they determined the velocity $V$ of the melt layer, when the melting of a solid overlain by a fluid of higher temperature leads to vigorous compositional convection. The timescale and wavelength for growth of the instabilities of the buoyant melt layer has been studied by a linear Rayleigh-Taylor stability analysis. The thickness of the melt layer, $h_{m}$ (comparable to the boundary layer thickness in our experiments) was given by~\cite{Kerr1994}: 
\begin{equation}
h_{m} \sim \left( \frac {PV\nu_{m}\rho_{m}} {g(\rho_{f}-\rho_{m})} \right) ^{1/2} 
\label{eq:Kerreq}
\end{equation}
where, $\nu_{m}$ is viscosity of the melt layer (in our case $\nu_{p}$), $\nu_{f}$ is the viscosity of the fluid (in our case $\nu_{a}$), $\rho_{m}$ and $\rho_{f}$ are the densities of the melt layer and fluid respectively (in our case $\rho_{p}$ and $\rho_{a}$ respectively), g is the acceleration due to gravity. The viscosity ratio is $\nu_{f} / \nu_{m}$ which in our case is $U$, $P$ is a function of $U$ ~(\cite{Kerr1994}) and $V$ is the velocity of the melt layer (in our case the through flow velocity across the mesh, $v_m$). In the inset of Figure~\ref{fig:spacingthickness}(b), we show the variation of $h_{m}$ with the viscosity ratio. 

For $U>1$, the plume spacing $\lambda$ is given by~\cite{Whitehead1975,Kerr1994,Jellinek1999}: 
\begin{equation}
\lambda \sim C h_{m} U^{1/3}
\label{eq:K1}
\end{equation}
Different predictions of the pre-factor $(C)$ on the right hand side of the above Eq~\ref{eq:K1} exist; ~\cite{Kerr1994} suggests a value of $C = \pi(2/3)^{1/3} \approx 2.74$, and~\cite{Whitehead1975} suggest that it is $C = 4.36$. In Figure~\ref{fig:spacingthickness}, we compare our results with  $\lambda$ obtained using the above Eq~\ref{eq:K1} with the different pre-factors for the right hand side.  
For $U<1$, the plume spacing $\lambda$ has been suggested to be~(\cite{Whitehead1975}): 
\begin{equation}
\lambda \sim 4.45 h_{m} U^{-1/5}
\label{eq:K2}
\end{equation}
In Figure~\ref{fig:spacingthickness}, we also compare our experimental results with the predictions from Eq~\ref{eq:K2}. Although we do not find a very good agreement with the predictions \tb{(Figure~\ref{fig:spacingthickness}(a))}, for $U>1$ the general trend of $\lambda$ increasing with $U$ 
is observed and for $U<1$, $\lambda$ \tb{increases with decrease in} $U$. 

In Figure~\ref{fig:spacingthickness}(b), we plot the plume spacing normalized by $h_{m}$ versus the viscosity ratio, $U$. \tb{The normalized plume spacing ($\lambda / h_{m}$) from our experiments show a weak dependence on $U$, with small variation in a narrow range between 10--30. }


\tb{In summary, we have compared our experimental results with theoretical predictions from high Rayleigh number convection: (i) over heated horizontal surfaces (\cite{Theerthan1998}), (ii) and driven by density differences across a mesh (\cite{Baburaj2005}), but these studies were conducted at $U = 1$ and they do not have an imposed through flow across the mesh, and their results will not be expected to hold for cases with viscosity contrast. We also compare theoretical results from melting driven by compositional convection including a viscosity contrast (\cite{Kerr1994}), but the exact details of the system is different from ours. Finally, we compare theoretical results from a Rayleigh-Taylor stability analysis, including viscosity contrast (\cite{Whitehead1975}), which may not be suitable for the convection regime in our experiments. All of these theories have been developed for similar physical systems, but no analytical work has been done to exactly capture the dynamics in our system. This gives rise to the inconsistency between the previous theories and our experiments. In future work, an analytical model for our system could be developed starting with the model given in \cite{Baburaj2008} (which is for $U = 1$ with through flow), and incorporating a viscosity contrast between the ambient and plume fluids. }
    
We have also visualised the planform plume structures at different heights from the mesh using a computer-controlled traverse setup.  This traverse enabled the controlled and precise movement of a horizontal laser sheet in the vertical direction. Videos (resolution: $1280\times720$ pixels, using a Nikon D90 camera) of the traverse experiments provide a detailed visualization of the plume structures at different horizontal cross-sections of the flow (selected montages and three-dimensional reconstructions are shown in supplementary material). 
    
\begin{figure*}
\centering
\includegraphics[width=1\textwidth]{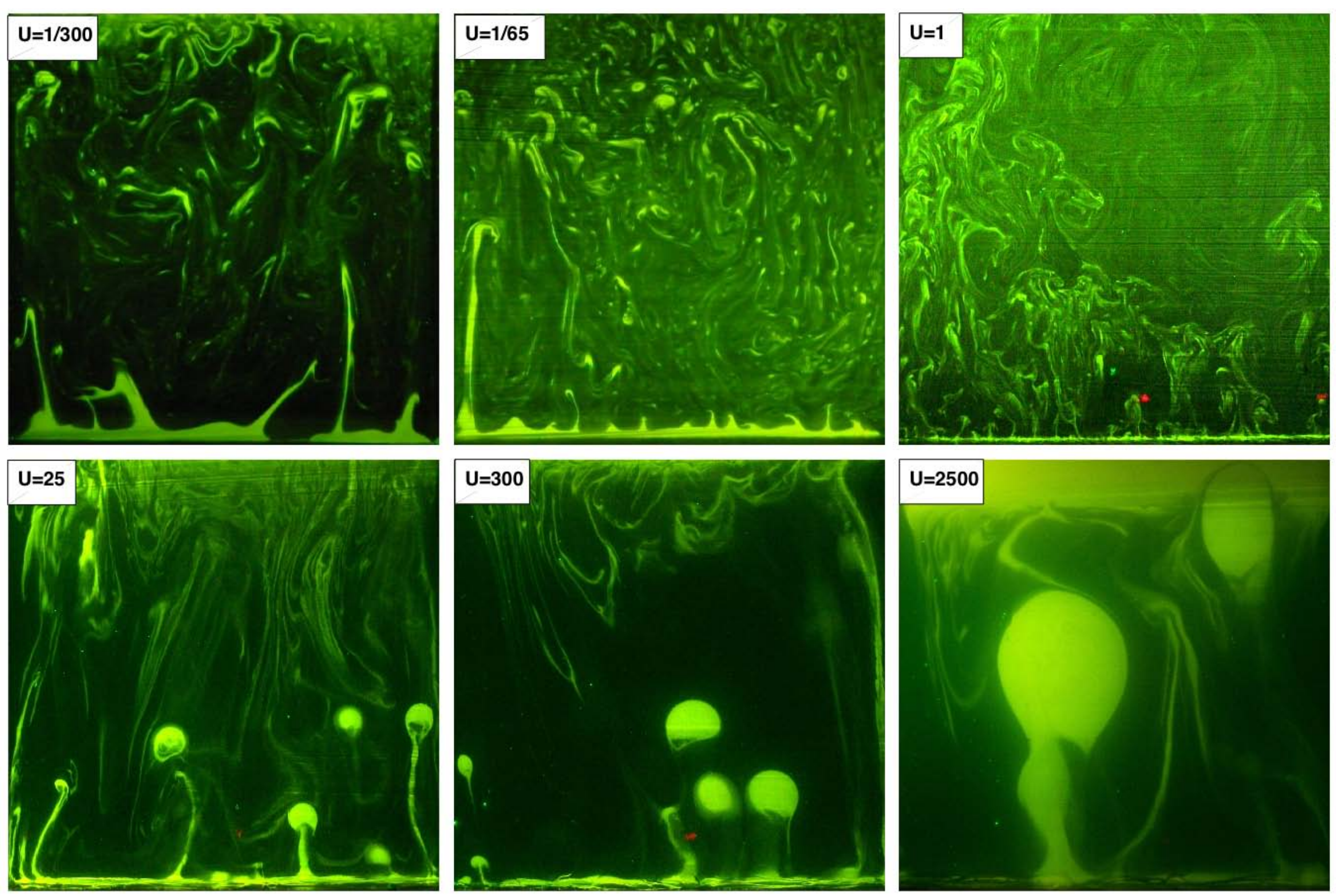}
\caption{Vertical section Images of rising plume structures. Images [1] to [6] correspond to viscosity ratio U = 1/300, 1/65, 1, 25, 300, 2500. The green line at the bottom is the boundary layer on the mesh and the plume rise direction is upwards. }
\label{fig:montagesideview}
\end{figure*}

\subsection{Plume Structure and dynamics in Vertical Sections}

In the vertical sections, the objective is to study the effect of $U$ on plume structure (size and shape) and plume rise velocities. Figure~\ref{fig:montagesideview} shows raw images of the vertical sections in different regimes of $U$ $(U = 1/300, 1/65, 1, 25, 300, 2500)$, and corresponding planforms are presented in Figure~\ref{fig:planformmontage}. The selection of these vertical sections (Figure~\ref{fig:montagesideview}) represents distinct regimes: (a) subduction regime ($U=1/300$ and $1/65$), (b) turbulent Rayleigh-Benard convection ($U=1$), (c) initiation of spherical-blob shaped plumes ($U=25$ and $300$) and (d) mature spherical-blob plume regime ($U=2500$) that corresponds to hot mantle plumes~(\cite{Olson1985}). In these experiments, the vertical laser sheet positioned in the middle of the test section illuminates the test-section, perpendicular to the mesh (see supplementary material for more detailed visualizations of the plume structure). In Figure~\ref{fig:montagesideview}, the green area just above the mesh, indicates the boundary layer developed over the mesh, \tb{and the images shown correspond to a time half-way through the experiment ($\sim$ 10 min)}. For the cases when $U$ is greater than one, the bottom boundary layer is thin, from which plumes emanate. However, as the value of $U$ is progressively reduced $(U=1/25$ and $1/300)$ thickness of the boundary layer increases.  When $U = 1$, the plumes are small with mushroom shaped heads, and plumes rapidly mix into the ambient fluid. We observe that as $U$ is progressively increased, from $U= 25$ onwards, plume head changes from mushroom-like structures to spherical-blobs, and for $U>300$ these blobs dominate the dynamics, as also observed in earlier studies~(\cite{Jellinek2002,Lenardic2009}). At higher values of $U$, plumes retain their identity for a longer distance and rise all the way to the top due to very little mixing of the plume and ambient fluids. The plume fluid rises and ponds to form a distinct layer of buoyant fluid at the top of the test section. The height of this ponded layer \tb{is time dependent} and is maximum when $U = 2500$ (Figure~\ref{fig:montagesideview}). Also in the case of higher $U$, a starting plume leaves behind a conduit of low viscosity fluid - such conduits have also been observed in the point-source experiments of~\cite{Olson1985}. Subsequent plumes that arise near the remnant conduits tend to flow through them as they offer a low resistance path. The plume heads also have a tendency to align themselves towards the conduits and may end up rising in an inclined trajectory~(\cite{Olson1985}). The plume head deforms when it interacts with a conduit and accelerates upwards. In cases where $U > 1$, the plume heads also interact with each other provided they are in sufficient proximity to each other. This `clustering' of plumes~(\cite{Manga1997,Kelly1997}) results in the merging of two or more starting plume heads into a single larger plume head. 

\begin{figure} 
\centering
\includegraphics[width=0.9\textwidth]{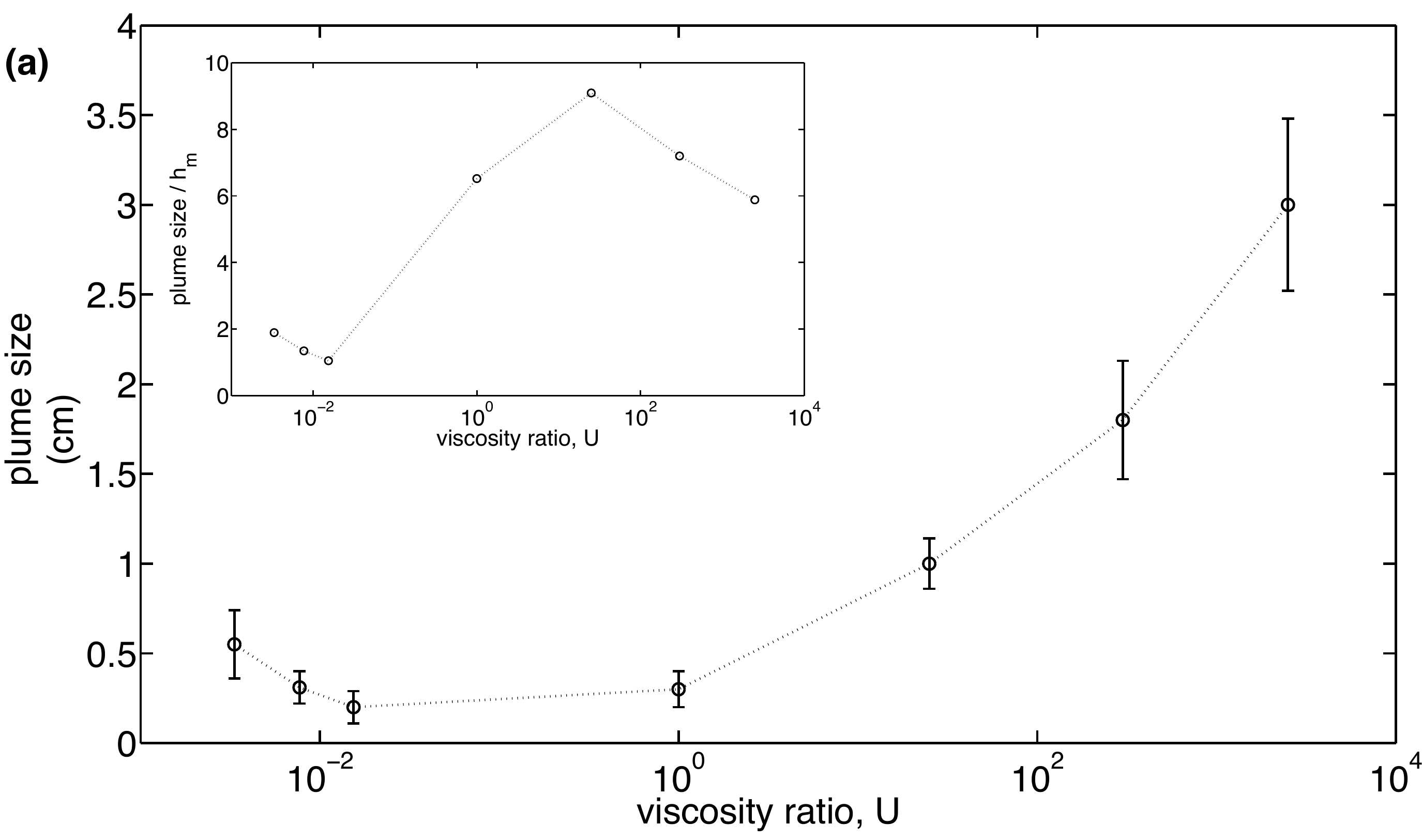}
\includegraphics[width=0.93\textwidth]{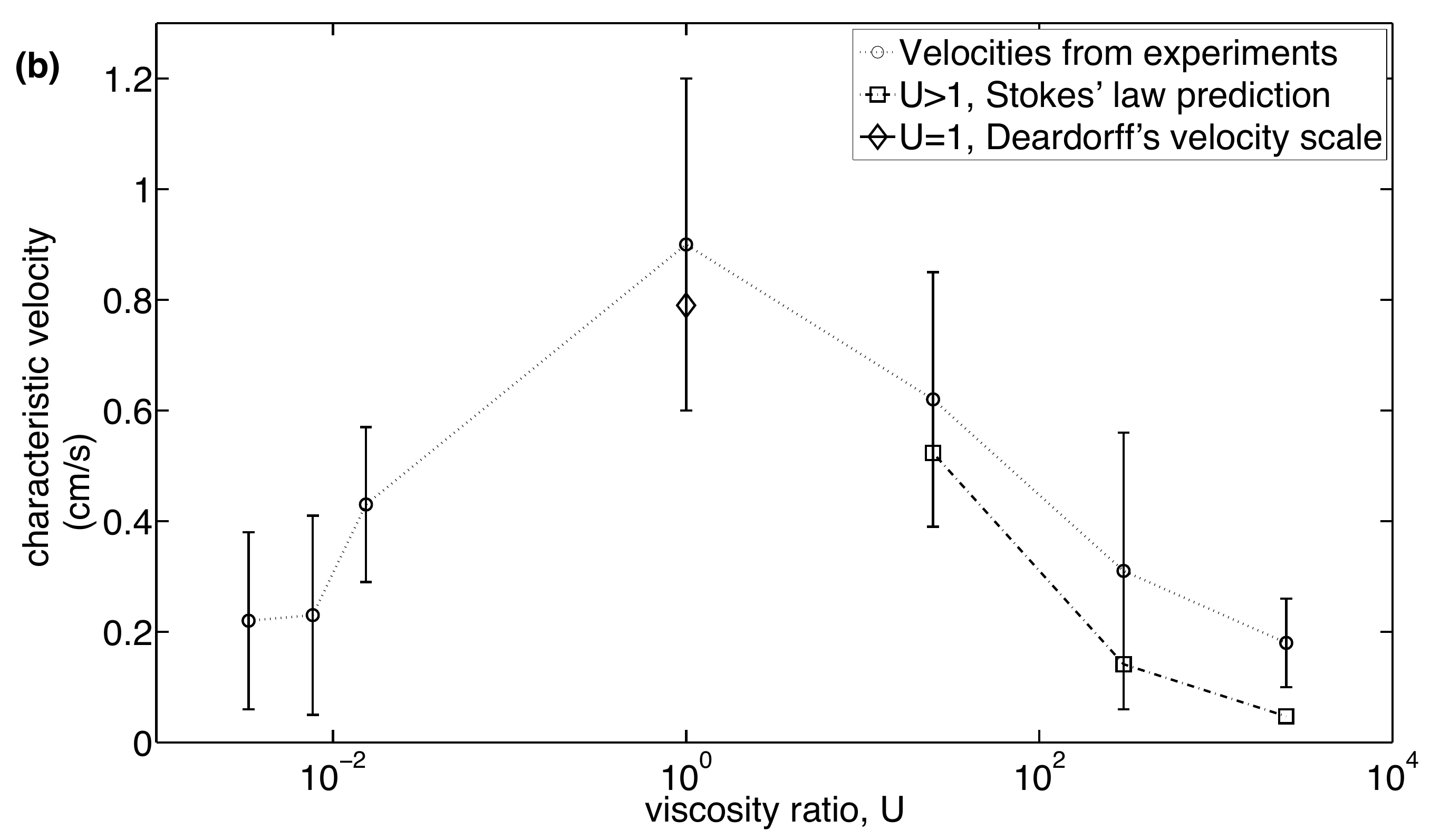}
\caption{Plots of: (a) Plume size (in cm) versus viscosity ratio, $U$. Inset: Normalized plume size (plume size / $h_{m}$ versus $U$. (b) Plume characteristic velocities (in $cm$ s$^{-1}$) versus viscosity ratio, $U$. }
\label{fig:sizevelvsU}
\end{figure}

Subduction of lithospheric plates into the earth's mantle is similar to cold, descending plumes in Rayleigh-Benard convection. Although there are different mechanisms suggested for subduction,  one of the accepted models is that subducting plates (like oceanic crusts) are denser~(\cite{Conrad2002,Richter1978,Hager1984,Kincaid1987,Stern2007,Schubertbook}) and more viscous than the mantle fluid into which they sink~(\cite{Kincaid1987}), and is analogous to $U < 1$ in our experiments. We observe that the plumes detach from the boundary and rise as broad sheets (slab-like) and penetrate deep into the test-section fluid (Figure~\ref{fig:montagesideview}, $U = 1/300, 1/65$). The head of these plumes are not bulbous, and their penetration into the ambient fluid is maximum for the smallest $U (= 1/300)$ and decrease with increase in $U$. Also, as stated earlier, the thickness of the boundary layer on the mesh, from which plumes emanate,  is largest for the smallest $U(= 1/300)$ and decreases with increase in $U$.

For a more quantitative analysis, individual frames were extracted from the videos of the vertical section experiments ($25 fps$, resolution: $704\times576$ pixels, $20$ minutes) and analyzed. In each experiment, the individual plumes were tracked in their corresponding frames to obtain their size and velocity information. The plume sizes were estimated knowing the test section length scale. The maximum equivalent diameter of a sample plume selected at mid height in the vertical section is referred to as the size of the plume. Measuring the plume size in the $U = 1$ experiment is difficult because of the small size and rapid mixing, the size is estimated to be $\sim0.3 cm$. For the other experiments at different $U$, we determine the characteristic plume size based on many sample measurements. Figure~\ref{fig:sizevelvsU}(a) shows a plot of characteristic plume size versus $U$ for the vertical section experiments. In Figure~\ref{fig:sizevelvsU}(a), we see little variation in plume size for $U < 1$; there is a slight increase for $U < 1/65$. However, the plume size increases steadily for $U > 1$. In the inset of Figure~\ref{fig:sizevelvsU}(a), we show the plume size normalized by $h_{m}$, this measures the relative length scale of the plume size and the boundary layer thickness. We see a weak dependence for $U < 1$, but when $U > 1$ the normalized plume size varies in the range from 6 to 9.  

The velocities of the rising plumes were estimated by tracking the vertical position of the plume-head over time. From this, we obtain the height versus time data for each plume \tb{(see supplementary information)}. We fit a polynomial onto this height versus time data and take the first derivative to obtain a range of plume rise velocities. We identify a characteristic velocity for each case of $U$ based on a number of measured samples. Measuring this characteristic velocity for the $U=1$ case was difficult because of the rapid mixing of the plumes, and the estimated value  of the velocity is $\sim0.9 cm s^{-1}$. Figure~\ref{fig:sizevelvsU}(b) shows a plot of characteristic plume velocity versus viscosity ratio $U$ for all the vertical section experiments. Figure~\ref{fig:sizevelvsU}(b) reveals that the characteristic plume rise velocities are highest for $U = 1$ and decrease with increasing $U$ ($U>1$) and also with decreasing $U$ ($U<1$). When $ U = 1$, the 
velocity is comparable to the relevant velocity scale in the bulk region of turbulent natural convection, given by the Deardorff velocity scale~(\cite{Deardorff1970}): $w = (g\frac {\Delta\rho} {\rho} v_{m}H)^{1/3}$ $\sim0.79 cms^{-1}$ at a height of $H = 2.5 cm$ (see Figure~\ref{fig:sizevelvsU}(b)).  When $U>1$, the plume rise velocity is comparable to the terminal velocity of a buoyant sphere rising in a viscous liquid, which is given by Stokes' law: $v_s = (1/18)(\Delta\rho/\mu)gD^2$, where, $\Delta\rho$ is the density difference between the particle and the fluid, $\mu$ is the absolute viscosity of the surrounding liquid, g is the acceleration due to gravity, and $D$ is the diameter of the sphere. In Figure~\ref{fig:sizevelvsU}(b) we compare our experimental results to the estimated Stokes rise velocities for the $U>1$ cases: $U = 25, 300, 2500$. We observe that the plume rise velocities in the present experiments are slightly higher than the Stokes' prediction. In our experiments, the rising low viscosity plumes leave conduits behind them. The plumes which rise subsequently near these conduits are surrounded by a thin layer of low viscosity fluid, and as a consequence, they reach a higher terminal velocity than that predicted by Stokes' law. Since our experiments cover three very different regimes ($U < 1$, $U = 1$, and $U > 1$), we do not have a common velocity scale to normalize the velocity variation versus $U$ in Figure~\ref{fig:sizevelvsU}(b).

\subsection{\label{sec:mixing}Mixing}

\begin{figure}
\centering
\includegraphics[width=0.4\textwidth]{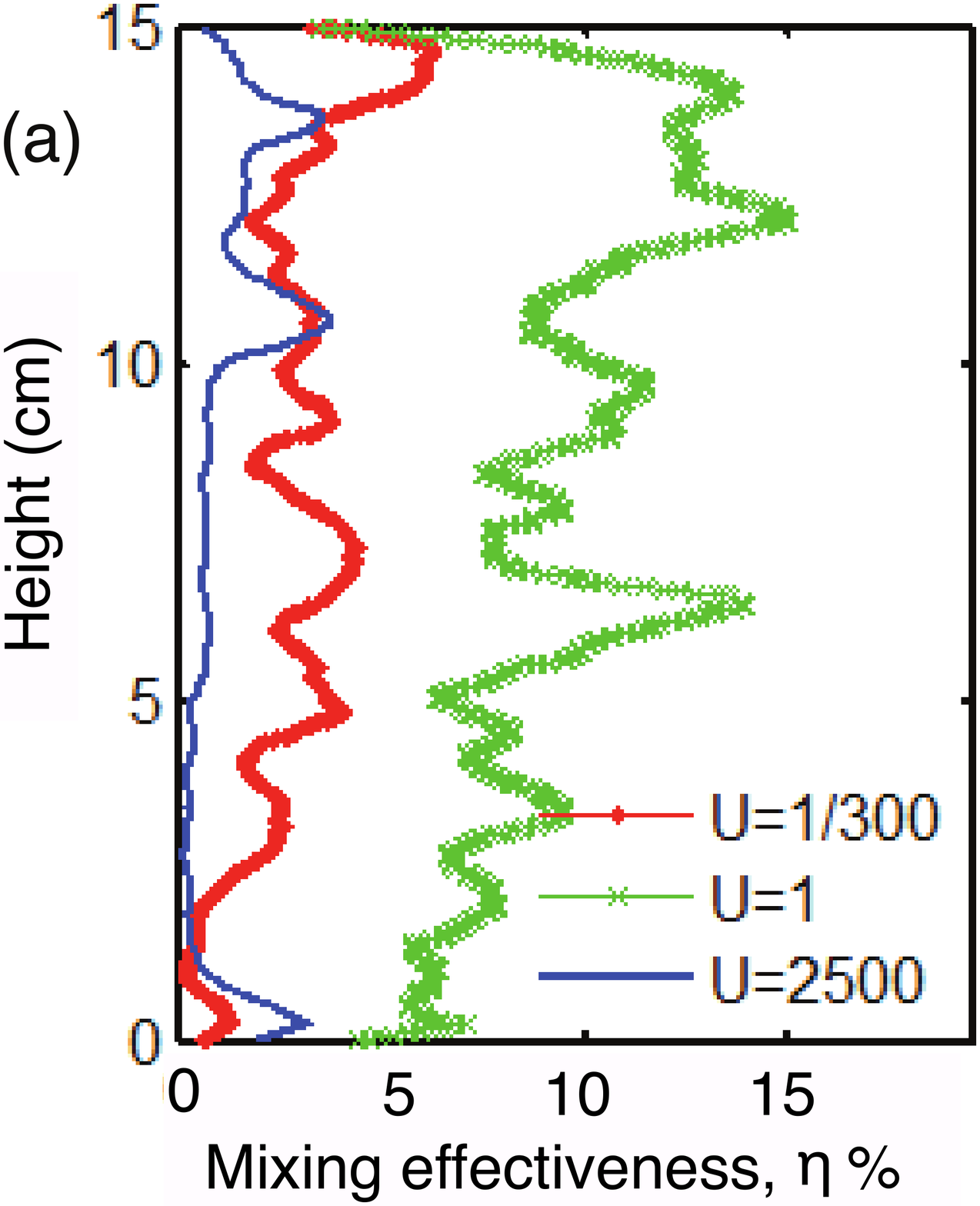}
\includegraphics[width=0.8\textwidth]{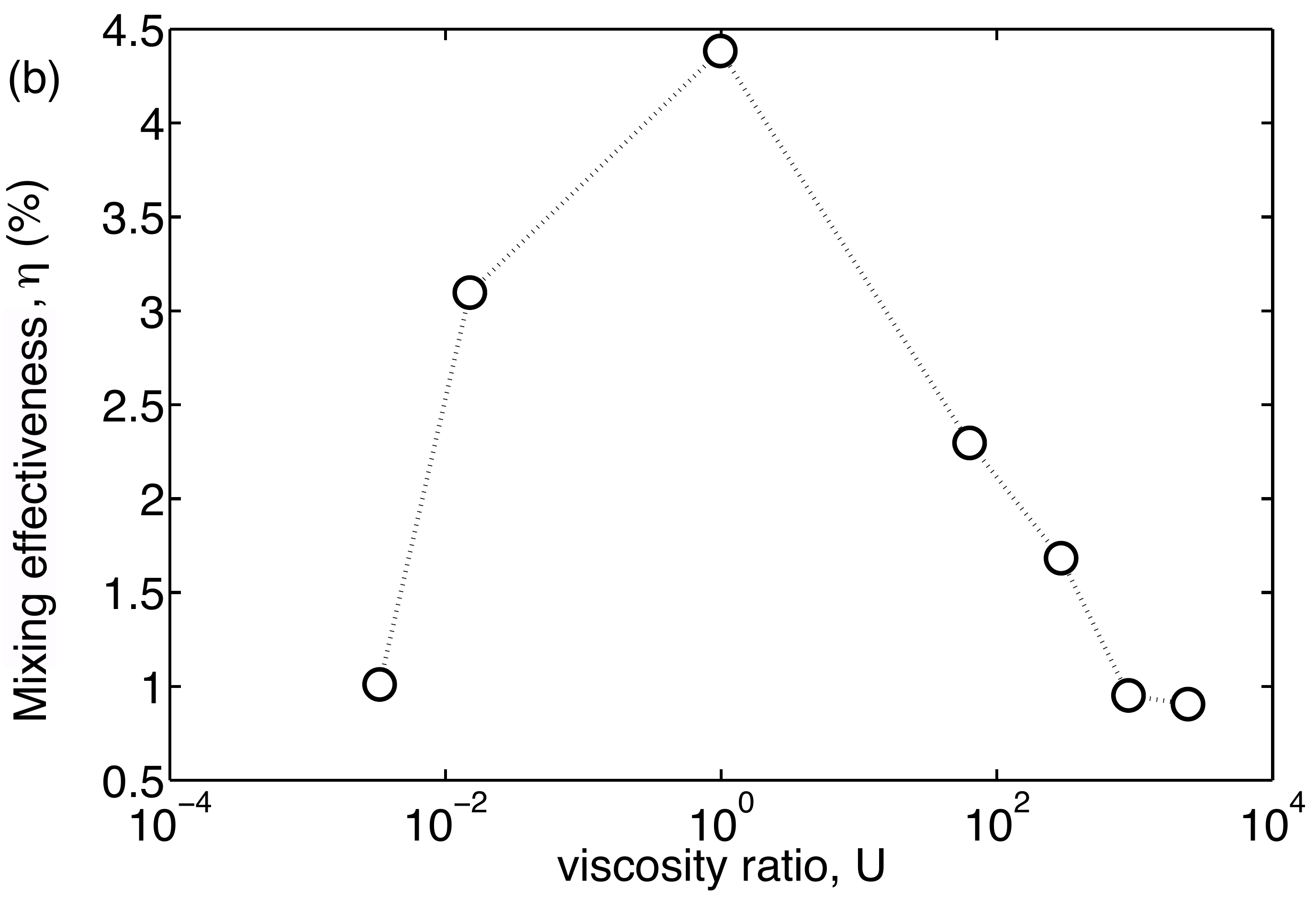}
\caption{Mixing effectiveness, $\eta$ of the plumes with the ambient fluid. (a) $\eta$ as a function of the height above the mesh, for the cases $U=1/300$, $U=1$ and $U=2500$ (corresponding to vertical section images [1], [3] and [6] in Figure~\ref{fig:montagesideview}). (b) $\eta$ versus the viscosity ratio, $U$ (for planform images at a height of 2 cm above the mesh).}
\label{fig:mixingvsH}
\end{figure}

The rate of mixing of the plumes with the ambient mantle fluid determines the longevity and the extent to which a plume rises (hot rising plume) or descends (cold subducting plate) in the mantle. Quantification of mixing is constrained in a laboratory model because of the fact that the total duration of the plume rise in these experiments is only of the order of few tens of seconds. During this short period, molecular diffusion can make the tracer dye to diffuse over a short length-scale of $1$-$2$ pixels ($n_{\delta}$). The mixing quantification involves two measures, (a) homogeneity of dye concentration, and (b) the length scale of segregation.  To quantify the mixing, we define a mixing effectiveness parameter, $\eta$ as: 
\begin{equation}
\eta = \left[ \frac {(\bar{I} - \sigma_I)} {\bar{I}} \right] \left[ \frac {N_{cr}} {N_{max} } \right]
\end{equation}
which is a product of above stated two measures. $\eta=1$ corresponds to complete mixing, and $\eta=0$ would represent no mixing. In $\eta$, the first term in the square bracket indicates the homogeneity in the dye concentration and is based on the standard deviation $(\sigma_I)$ and the mean value of the dye concentration $(\bar{I})$. The second square bracketed term indicates the length scale of segregation (segregation-index) and is the ratio between number of times the intensity profile crosses the mean value $(N_{cr})$ to the maximum possible value for this cross-over $(N_{max})$. For a given resolution of the image, $N_{max}$ corresponds to the ratio of total number of pixels ($N$) along a line to the number of pixels representing the diffusion length scale, $n_{\delta}$, which in the present study is taken as one. Thus by computing $\eta$ we can quantify effective mixing in the flow, which is an indication of the longevity and extent of plume penetration 
as a function of viscosity ratio, $U$. If the standard deviation $(\sigma_I)$ is small, then the fluid more or less has a uniform dye concentration and hence it is better mixed. Whereas $\sigma_I \to \bar{I}$, i.e. in the limit when the standard deviation is of the same magnitude as the mean value of the dye concentration, then the dye concentration in the fluid has large fluctuations and hence is poorly mixed. Similarly, if the dye concentration variation along a line in the image crosses the mean dye concentration value a large number of times, then length scale over which the fluctuation is occurring is small and this indicates a better mixing. In the opposite limit, when the number of crossings is small, the fluid has large-scale segregation and hence mixing is poor. Thus, lower values of $\eta$ indicate poor mixing (either due to high fluctuation in dye concentration or due to large length scale segregation) and higher values of $\eta$ indicate good mixing (in this case, \textit{both} the magnitude of concentration fluctuations and the length scale of segregation are small). The performance of the mixing effectiveness parameter, $\eta$, has been tested using synthetic images (see supplementary material).

The mixing analysis using the above definition of mixing effectiveness $(\eta)$ is carried out on a set of vertical section images of the flow field from the present experiments. 
The vertical mixing profile is shown in Figure~\ref{fig:mixingvsH}(a), for the cases $U=1/300$, $U=1$ and $U=2500$ (corresponding to images [1], [3] and [6] in Figure~\ref{fig:montagesideview}). The mixing effectiveness, $\eta$ is highest in the case where $U=1$ and is lowest when $U=2500$. Excluding the upper most region (where the plume fluid is pooled during the experiment), the mixing effectiveness increases with height.

The results obtained from the mixing analysis of the planform images at different values of $U$, at a height of $2$ cm above the mesh, are presented in Figure~\ref{fig:mixingvsH}(b). We select this height due to two reasons: (i) for $U=1$, at heights very close to the mesh, since the mixing is very efficient we will find almost a homogenous profile over time, and (ii) for the $U<1$ cases, it is necessary to be sufficiently far away from the boundary layer which can be quite thick ($\sim4$ mm). In Figure~\ref{fig:mixingvsH}(b), We see that the mixing effectiveness is best at $U = 1$ and decreases either if $U$ increases ($U>1$) or if $U$ decreases ($U<1$). 

\tb{The images used for mixing quantification in Figure~\ref{fig:mixingvsH}(a) are from vertical section experiments and in Figure~\ref{fig:mixingvsH}(b) the images are from planform experiments at different heights. Although the overall mixing trends between the different cases of $U$ are captured in both figures, the spatial mixing profiles in the two perpendicular planes can be slightly different. Also, since the mixing is a complex time-dependent process, and since our quantification is based on single images, fluctuations are expected due to plume dynamics over time, and hence the $\eta$ values in Figure~\ref{fig:mixingvsH}(a) and (b) are slightly different.}

Our results on mixing in the $U<1$ regime is in contrast to those reported by~\cite{Jellinek1999}. In their study, the \textit{mixing efficiency} reflects the available potential energy of the rising buoyant fluid parcels, whereas our interest is to study the longevity of the plumes by tracking the identity of the plume fluid. Hence, our definition of mixing effectiveness is based on the differences in dye concentration at a given cross-section, also taking account into the segregation. We also measure the mixing effectiveness during the experiment, while~\cite{Jellinek1999} withdraw samples at the end of the experiment to measure the mixing efficiency. \cite{Jellinek1999} found that the mixing efficiency for cases with $U < 1$ is almost as good as the $U = 1$ case. Thus, the definition of mixing needs to be interpreted very carefully to compare results between \cite{Jellinek1999} and the present experiments. In summary, our results suggest that the identity of hot plumes is maintained over its passage in the mantle, chiefly due to the viscosity contrast that results in poor mixing.

 Here it is interesting to note the Reynolds numbers ($Re_a$) for the convecting layer. The $Re_a$ is small ($\sim O(1)$) when the viscosity of the ambient fluid is high (Experiment sets 10-12) $U>1$ and the observed mixing is low. Even when $Re_a$ is high (based on ambient viscosity) in experiments with $U<1$  (sets 13-15), the observed mixing is low. Here, however, it is to be noted that the plume-viscosity is high and the Reynolds number for the plume is low, thus the low mixing observed in these cases is also understandable.  

\subsection{Effect of varying the buoyancy flux} \label{sec:varyingflux}

\begin{figure}
\centering
\includegraphics[width=0.6\textwidth]{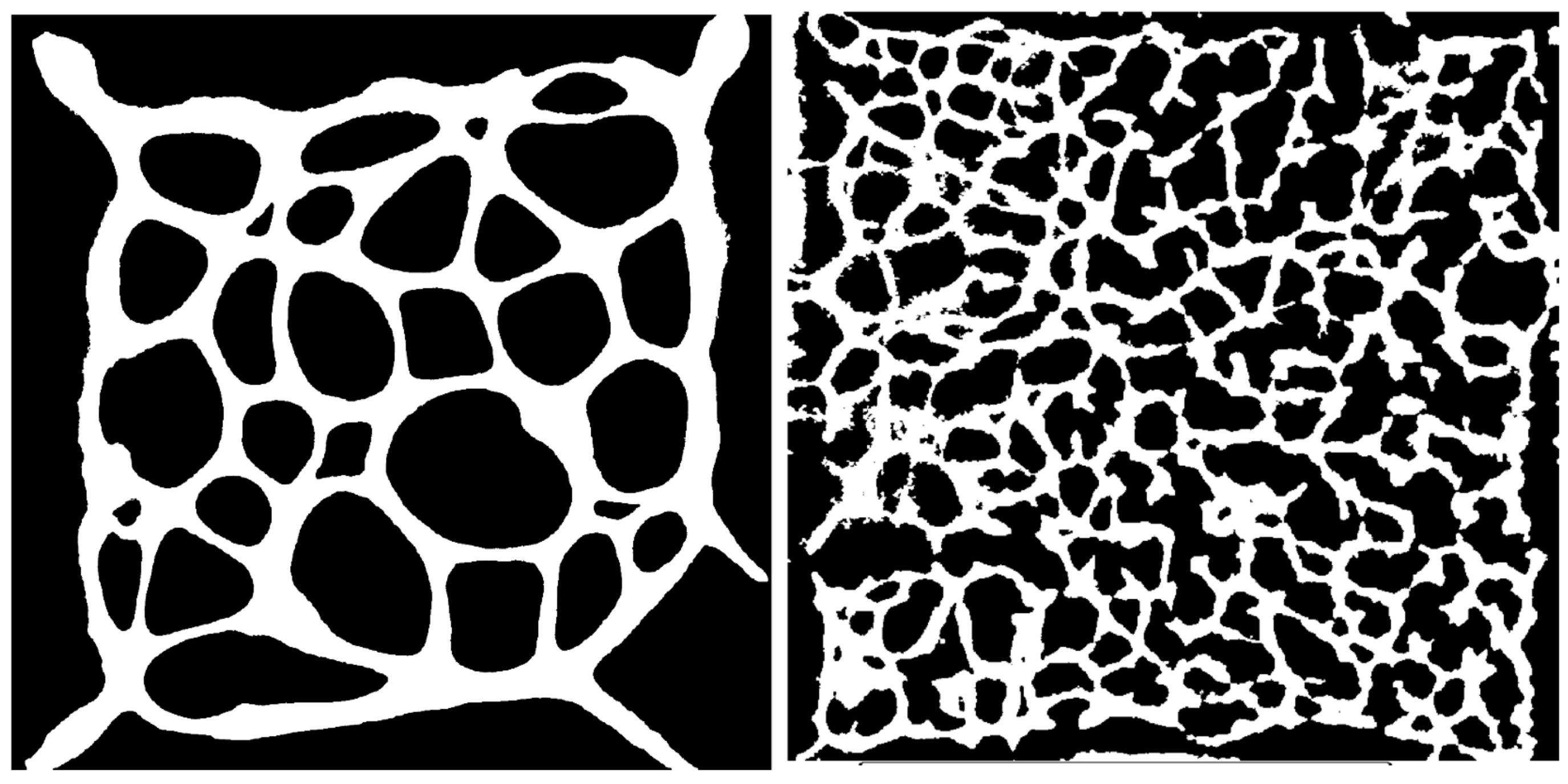}
\caption{Near-wall plan view binary images of plume structures in Experiment sets $13, 21$ $(U = 1/300)$. The structures in white color correspond to the plumes and the background is black. In the left image, the initial $\Delta\rho$ = 2.4 kg $m^{-3}$ and in the right image, the initial $\Delta\rho$ = 4.8 kg $m^{-3}$  (doubling $\Delta\rho$ )}
\label{fig:doublerho}
\end{figure}

Here, we report preliminary results from experiment sets 17--20 and 21 (Table~\ref{tab:expparam}), where the buoyancy flux, $B$ was varied (by changing $v_m$ or $\Delta \rho$) and the viscosity ratio $U$ was held constant. In experiment set 21, we studied the effect of doubling the density difference between the two fluids keeping the same viscosity ratio $U (=1/300)$ and $v_m$, as in experiment set 13. The binary images of the planforms are shown in Figure~\ref{fig:doublerho}, and it is observed that doubling the density difference driving the convection (2$\Delta \rho$), decreases the plume spacing. This is shown quantitatively in Figure~\ref{fig:spacingthickness}, where the plume spacing decreases at the same U (=1/300) (shown by solid and open black circles). The decrease in plume spacing on doubling the density difference is qualitatively consistent with theory~(\cite{Kerr1994}); Eq~\ref{eq:Kerreq} reveals the same trend. 
This decrease in spacing is also in conformity with other earlier results~(\cite{Baburaj2005,Theerthan1998}), see Eqs~\ref{eq:Theerthan} and \ref{eq:Baburaj}, where increasing the density difference or temperature difference resulted in closely packed plumes. \tb{A recent numerical study of geometrical confinement in Rayleigh-Benard convection~(\cite{Chong2015}) also found a decrease in plume spacing with increase in Rayleigh number.}

\begin{figure}
\centering
\includegraphics[width=0.6\textwidth]{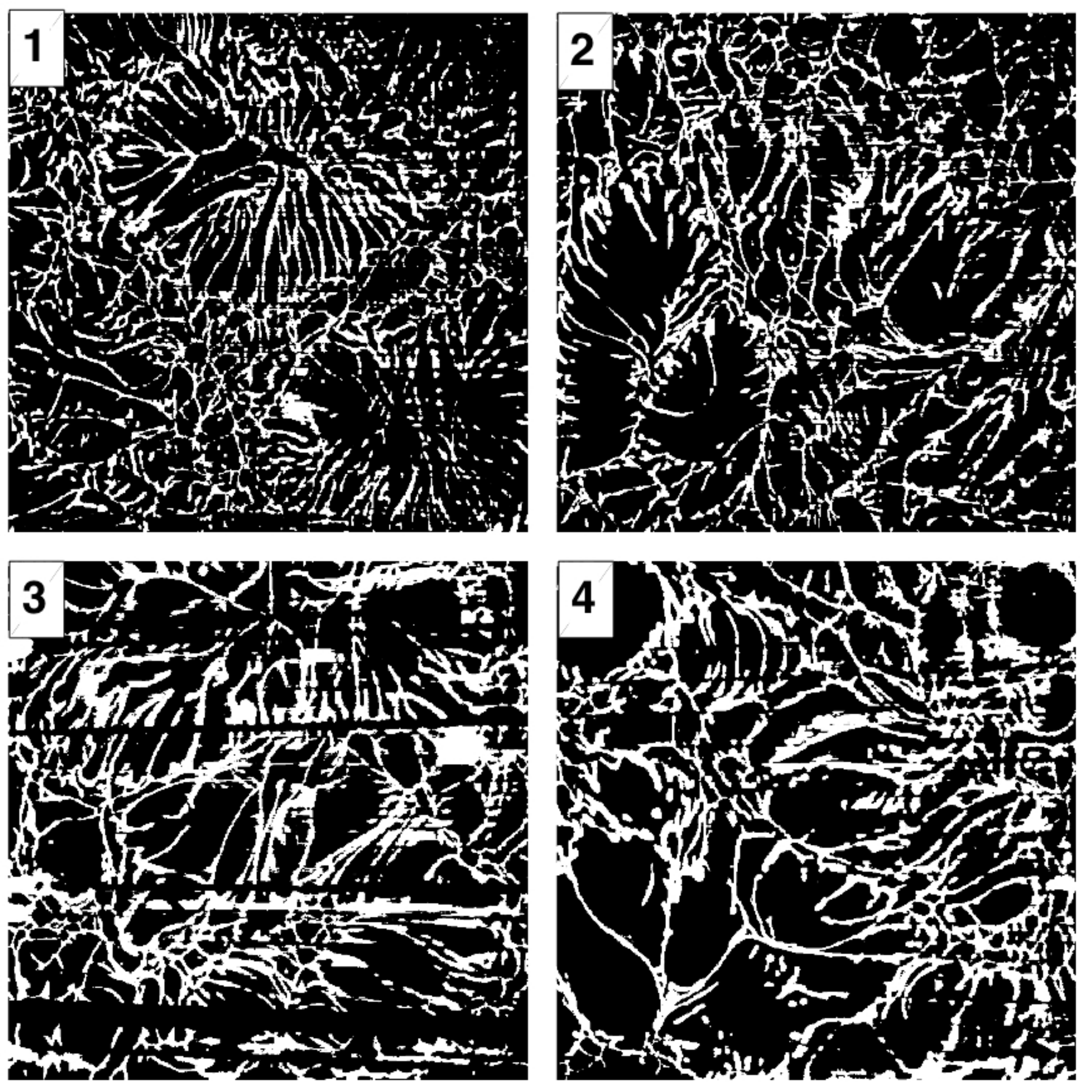}
\caption{Near-wall plan view images (binary) of plume structures in Experiment sets $17-20$ $(U = 1)$. The structures in white color correspond to the plumes and the background is black. Images [1] to [4] correspond to experiments where the through flow velocities across the mesh, $v_m$ = 0.020, 0.041, 0.062, $\&$ 0.083 cm $s^{-1}$ respectively.}
\label{fig:varyingQ}
\end{figure}

In the present experiments, an alternate way of varying the buoyancy flux is by changing the mesh through flow velocity (see Eq~\ref{eq:fluxeqn1}). In experiment sets 17--20 (Table~\ref{tab:expparam}), the viscosity ratio, $U = 1$, the density ratio was maintained constant ($\Delta\rho$ = 2.4 kg $m^{-3}$), and the through flow velocity, $v_m$ was varied to study its effect on the near-wall planform structure; $v_m$ = 0.020, 0.041, 0.062 cm $s^{-1}$ $\&$ 0.083 cm $s^{-1}$ (base case for the constant buoyancy flux experiments).  Figure~\ref{fig:varyingQ} shows binary images of the planform structures, in the experiments with different $v_m$. It is seen that the plume spacing is minimum at the lowest $v_m$ (0.020 cm $s^{-1}$) and is maximum at the highest $v_m$ (0.083 cm $s^{-1}$). This change in plume spacing with $v_m$ at a constant U (=1) is shown quantitatively in Figure~\ref{fig:spacingthickness} (shown by `x' symbols). Once again, Eq~\ref{eq:Kerreq}~(\cite{Kerr1994}) captures this trend, as spacing $\lambda$ is proportional to $h_m$, which in turn is proportional to $\sqrt{v_m}$. 
Referring back to Eq~\ref{eq:fluxeqn1}, the buoyancy flux, $B$, has two components: the advective and the diffusive components. In the limiting case of $v_m\rightarrow0$, the diffusive component becomes dominant and determines the plume spacing~(\cite{Baburaj2005,Baburaj2008}).    



In summary, separating the effect of density and mesh through flow velocity has a contrasting effect on the plume spacing, and this behaviour is well captured by the expression (Eq~\ref{eq:Kerreq}) suggested by~\cite{Kerr1994}. The increase in plume spacing with increasing through flow velocity is an interesting result which warrants a detailed further investigation. 

\section{Conclusion} \label{sec:con}

\begin{figure}
\includegraphics[width=1.1\textwidth]{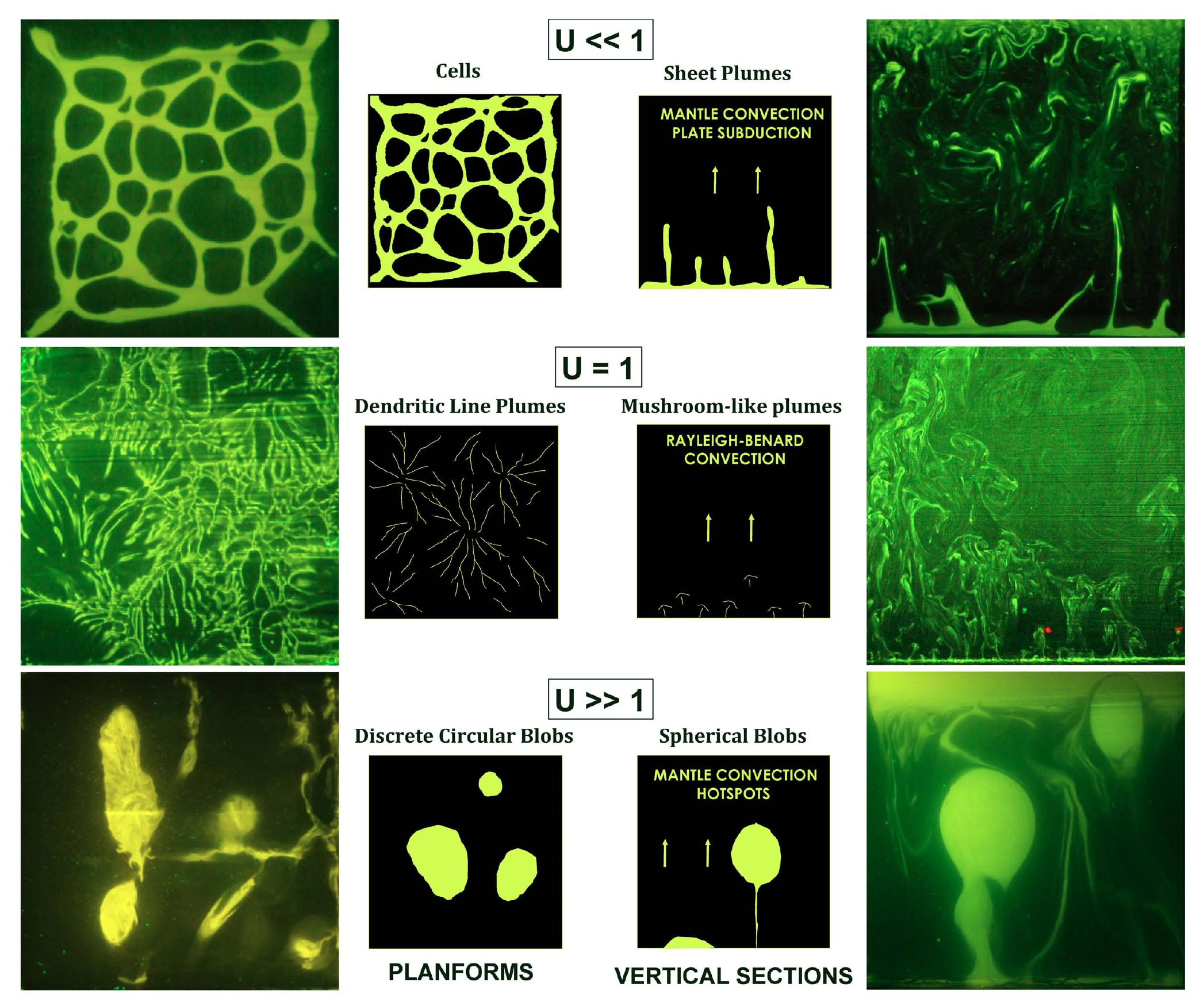}
\caption{The characteristic plume structures are shown at the different regimes of viscosity ratio, $U$ (viscosity of ambient fluid / viscosity of plume fluid) relevant in mantle convection. The near wall planform plume structures are shown on the left and the vertical section views are shown on the right. The arrows indicate direction of flow.}
\label{fig:APS_gallery}
\end{figure}

The viscosity ratio between the ambient mantle fluid and the plume, $U$, and high Prandtl numbers, are characteristics of mantle convection that sets it apart from the standard Rayleigh-Benard convection. Mantle convection is complicated due to variations in properties like pressure, temperature, rheology, fractionation, non-Boussinesq effects etc. In this paper we have presented results from experiments that capture the effects of variation in viscosity ratio on the plume size, spacing and mixing dynamics. In the experiments, the convection is driven by compositional buoyancy and is in the regime of high Schmidt numbers ($10-10^3$), high compositional Rayleigh numbers ($10^8-10^{11}$), high flux Rayleigh numbers ($10^{13}-10^{16}$), moderate Reynolds numbers ($1-10^3$) and viscosity ratios ($0.03 - 2500$). These regimes are very relevant to mantle convection, where the corresponding numbers (from~\cite{Jellinek1999}) are: Prandtl number $\sim10^{24}$, thermal Rayleigh number $\sim10^{6}$, thermal flux Rayleigh number $\sim10^{11}$, Reynolds number $\sim10^{-3}$, and viscosity ratio $\sim10^2$. We have presented flow visualizations of the plume structures using LIF, both in the planform and vertical sections. The buoyant plumes exhibit diversity in their morphology, thickness and spacing depending on the viscosity ratio (U) between the ambient and plume fluids.

The observed morphological changes have been summarized in Figure~\ref{fig:APS_gallery}. Here, the left most column shows the planform views and the right most column shows the vertical sectional views. The middle two columns in Figure~\ref{fig:APS_gallery} show cartoon representations of these results. The selection of $U$ values for which the results have been presented, corresponds to three distinct regimes, namely subduction $(U\ll1)$, Rayleigh-Benard type convection in the high
Rayleigh number regime $(U=1)$ and hot mantle plumes $(U\gg1)$. The results in Figure~\ref{fig:APS_gallery} indicate that in the subduction regime $(U\ll1)$, the near wall structure consists of a cellular pattern formed over a thick boundary layer from which unstable, high viscosity fluid rises as sheet plumes. These sheet plumes are morphologically similar to subducting crustal-slabs moving into the
mantle. When $U=1$, a standard dendritic structure of line-plumes are observed, with mushroom-heads. $U\gg1$ corresponds to the hot-mantle plumes - here we observe spherical intermittent plumes as reported by~\cite{Olson1985}. These plumes at times pass through low resistance conduits, left by earlier plumes. Also, we observe that nearby plumes merge to form super-plumes. On both sides of $U=1$, (either increasing $U$ above 1, 
or decreasing $U$ below 1), the characteristic spacing and plume thickness increases. 
However, in these regimes (either when $U > 1$ or $U < 1$) the mixing effectiveness decreases on both sides of $U = 1$, where it is maximum. The best mixing occurs when there is no viscosity contrast between the ambient and plume fluids $(U=1)$.
Earlier investigators~(\cite{Campbell1985,Turnerbook,Mani2006}) have reported reduced mixing, even with high plume Reynolds number, when the ambient viscosity is higher than the plume fluid. 
The reduced mixing makes it possible for hot-plumes to rise to the top, maintaining their identity. Thus, the results presented here capture some of the salient features of mantle convection. Quantitative results on characteristic length scales of plumes have been presented and compared with earlier studies. 
\tb{The non-dimensional plume spacing from our experimental data shows a small variation over a large viscosity range, and does not show an agreement with previous theories of \cite{Kerr1994} and \cite{Whitehead1975}.}
The viscosity contrast ratio $U$ turns out be an important factor in  determining mixing, plume spacing and its longevity; and also by varying $U$, one could capture observed mantle convection features, namely hot-spots and subduction regimes in a laboratory experiment.  More studies are required to establish the effects of through flow velocity on the plume structure and mixing.

\section{Acknowledgments}
We thank Baburaj A. P. for useful discussions, G. U. Kulkarni and Radha for the mesh imaging measurements, V. Kumaran and Madhusudan for the viscosity measurements. We also acknowledge financial support from the Jawaharlal Nehru Centre for Advanced Scientific Research.

\section{References}









\bibliographystyle{elsarticle-harv.bst}

\end{document}